\documentclass[aps,prb,10pt,twocolumn,letterpaper,amsmath,amssymb]{revtex4-2}

\usepackage[utf8]{inputenc}
\usepackage{graphicx}
\usepackage{subfigure}
\usepackage[usenames,dvipsnames]{xcolor}
\usepackage{epstopdf}
\usepackage{amsmath,amsthm,amssymb}
\usepackage{tcolorbox}
\usepackage{latexsym}
\usepackage{bm}
\usepackage{dcolumn}
\usepackage{braket}
\usepackage[normalem]{ulem}
\usepackage{times}
\usepackage{hyperref}
\usepackage{enumitem}

\newcommand{\ignore}[1]{}
\newcommand{\red}[1]{\textcolor{black}{#1}}
\newcommand{\rred}[1]{\textcolor{black}{#1}}
\newcommand{\rrred}[1]{\textcolor{black}{#1}}
\newcommand{\rrrred}[1]{\textcolor{black}{#1}}
\newcommand{\blue}[1]{\textcolor{black}{#1}}

\newcommand{\qedblack}{\hfill $\blacksquare$}
\newcommand{\boxedp}{\boxed{+}}
\newcommand{\boxedm}{\boxed{-}}

\newcommand{\typea}{type-\emph{a}~}
\newcommand{\typeb}{type-\emph{b}~}

\begin{document}


\title{Site-decorated model for unconventional frustrated magnets:\\ Ultranarrow phase crossover and two-dimensional spin reversal transition}
\author{Weiguo Yin}
\email{wyin@bnl.gov}
\affiliation{Condensed Matter Physics and Materials Science Division,
Brookhaven National Laboratory, Upton, New York 11973, USA}

\begin{abstract}
The site-decorated Ising model is introduced to advance the understanding and experimental realization of the recently discovered one-dimensional (1D) finite-temperature ultranarrow phase crossover in an external magnetic field, while mitigating the geometric complexities of traditional bond-decorated models. \rred{The unconventional frustration and physics are clarified by exactly mapping the 1D site-decorated Ising model in a magnetic field onto a zero-field bond-decorated $J_1$–$J_2$ Ising model with conventional geometrical frustration.} Furthermore, although higher-dimensional Ising models in an external field remain unsolved \rred{exactly}, an exact solution for a spin-reversal transition---driven by an exotic, hidden `half-ice, half-fire’ state induced by site decoration---is derived. This transition, triggered by a slight variation in temperature or magnetic field\rred{---without changing its direction---}even in the weak-field limit, offers a promising route toward energy-efficient applications such as data storage and processing. The results \rrrred{suggest that site decoration offers} an avenue for materials and device design, particularly in systems such as mixed 
$d$-$f$ compounds, optical lattices, and neural networks, \rrrred{calling for further studies with site-decorated Heisenberg models}. \red{In addition, the site-decorated model offers a rigorous test ground for artificial intelligence (AI) in science, as the analytic derivation of the present results was not only validated but also improved by \rrred{a general-purpose large language model}, 
inspiring the use of AI as scientific discoverer.}
\\

\noindent DOI: \href{https://doi.org/10.1103/v5qd-5n7z}{10.1103/v5qd-5n7z}

\end{abstract}

{
}
\date{Received 11 February 2025; revised 6 February 2026; accepted 19 February 2026; published 10 March 2026}

\maketitle


\section{Introduction}

Finding states with desirable physical properties and energy-efficient phase transitions is a central challenge in various research fields, including condensed matter physics, materials science, quantum information, and microelectronics~\cite{Kivelson_24_book_statistical}. For example, magnetization reversal in ferromagnets (FMs) and ferrimagnets is crucial for data storage and processing, yet traditional field-driven spin flipping suffers from considerable coercive forces~\cite{Kim_NM_22_review_ferri}, while current-induced spin–orbit torque has emerged as a more effective alternative~\cite{Ji_24_spin-orbit-torque_review}. \red{Recently, fast spin reversal driven by a hidden half-ice, half-fire state at finite temperature was predicted to emerge in one-dimensional (1D) Ising ferrimagnets~\cite{Yin_Ising_III_PRL}.  This finding was unexpected---1D Ising and 1D/two-dimensional (2D) Heisenberg models with short-range interactions are well known to lack finite-temperature phase transitions~\cite{Mattis_book_08_SMMS,Ising1925,Mermin_PRL_theorem} and hence have been largely overlooked for their potentials in technological applications---and has been regarded as a guide to materials design~\cite{Ramirez_25_SmMn2Ge2}. It is thus interesting to get more insight into the fire-ice mechanism, including its applicability in higher dimensions, and make the connections to applications more versatile.}

\red{In a broader sense, we deal with} frustrated magnets, which are known to host a rich variety of unconventional states. 
including spin ice, spin glass, spin liquid, spin vortex, and skyrmions~\cite{NNano_13_review_skyrmion,Balents_nature_frustration,Miyashita_10_review_frustration}. Traditionally, frustration---termed geometric frustration---arises from competing spin-spin interactions, leading to macroscopic ground-state degeneracy~\cite{Ramirez_review_frustrated_magnets_94}. A well-established approach to engineering such frustration is bond decoration~\rrred{\cite{Balents_nature_frustration,Miyashita_10_review_frustration,Ramirez_review_frustrated_magnets_94,Fisher_PR_59_Ising_transform,Naya_54_Ising_honeycomb_kagome_magnetization,Syozi_55_Ising_Ferri_2D,Syozi_Ising_2D,Stephenson_CanJP_70_J1-J2-Ising-chain}}
, where additional spins are introduced along a bond \red{such that the frustration function (defined as the product of the spin-spin interactions along any closed contour of the original and additional bonds) has a negative sign~\cite{toulouse1977frustration}}. 
Finite-temperature phase transitions can be engineered by first computing the zero-temperature phase diagram, in which two phases with frustration-induced macroscopically different degeneracy have their own stable regions, and then placing the system near their phase boundary and on the side with less entropy. When heated up, the system will enter the other phase with lower free energy gained from high entropy~\cite{Miyashita_10_review_frustration}. For certain decorated 1D Ising models, this approach yielded anomalous phase-transition-like behaviors---termed pseudotransition---at finite temperature $T_0$ with a narrow transition width $2\delta T$ (note that $\delta T=0$ corresponds to a genuine phase transition)~\cite{005_Galisova_PRE_15_double-tetrahedral-chain,007_Torrico_PRA_16_Ising-XYZ-diamond-chain,009_review_Souza_SSC_18_Ising-XYZ-diamond-chain_double-tetrahedral-chain-spin-electron,010_Carvalho_JMMM_18_Ising-XYZ-diamond-chain_quantum-entanglement,011_Rojas_BJP_20_Ising-Heisenberg-tetrahedral_diamond,013_Rojas_PRE_19_previous_4_models,014_Rojas_JPC_20_Ising-Heisenberg_spin-1-double-tetrahedral-chain,015_Strecka_APPA_20_Ising-diamond-chain,015-7_Canova_CzechoslovakJP_04_Ising-Heisenberg_diamond_chain,015-8_Canova_JPC_06_Ising-Heisenberg-spin-S-diamond-chain,016_Strecka_book_chapter,017_Krokhmalskii_PA_21_3-previous-chains_effective_model}.
$2\delta T$ can be reduced by placing the system closer to the phase boundary; however, since genuine phase transitions in the 1D models occur only at zero temperature, $T_0$ approaches zero as $2\delta T \to 0$~\cite{017_Krokhmalskii_PA_21_3-previous-chains_effective_model}, undesirable for applications. Then we asked the following question: Can $T_0$ and $2\delta T$ be made independent? More specifically, can $2\delta T$ be reduced exponentially toward zero at fixed finite $T_0$? The phenomenon with this highly desirable feature, termed ultranarrow phase crossover (UNPC), was recently discovered \rred{to occur spontaneously---in the absence of an external magnetic field---}in 1D Ising ladders with bond-decorated rungs, \rred{unveiling an unconventional order parameter (OP) in which $T_0$ and $2\delta T$ are controlled independently by various interactions and decorations~\cite{Yin_MPT,Yin_icecreamcone}. Exploiting such a mathematical structure of OP, we found infinitely many spontaneous UNPC cases}~\cite{Yin_MPT,Yin_icecreamcone} but also established a paradigm of UNPC in bond-decorated single-chain Ising models in the presence of an external magnetic field~\cite{Yin_MPT_chain}. While finite-temperature phase transitions have been theoretically forbidden in this fundamental 1D statistical mechanics system for a century~\cite{Kivelson_24_book_statistical,Mattis_book_08_SMMS,Ising1925}, UNPC now enables them to be approached arbitrarily closely. 

\begin{figure}[t]
    \begin{center}
\includegraphics[width=0.99\columnwidth,clip=true,angle=0]{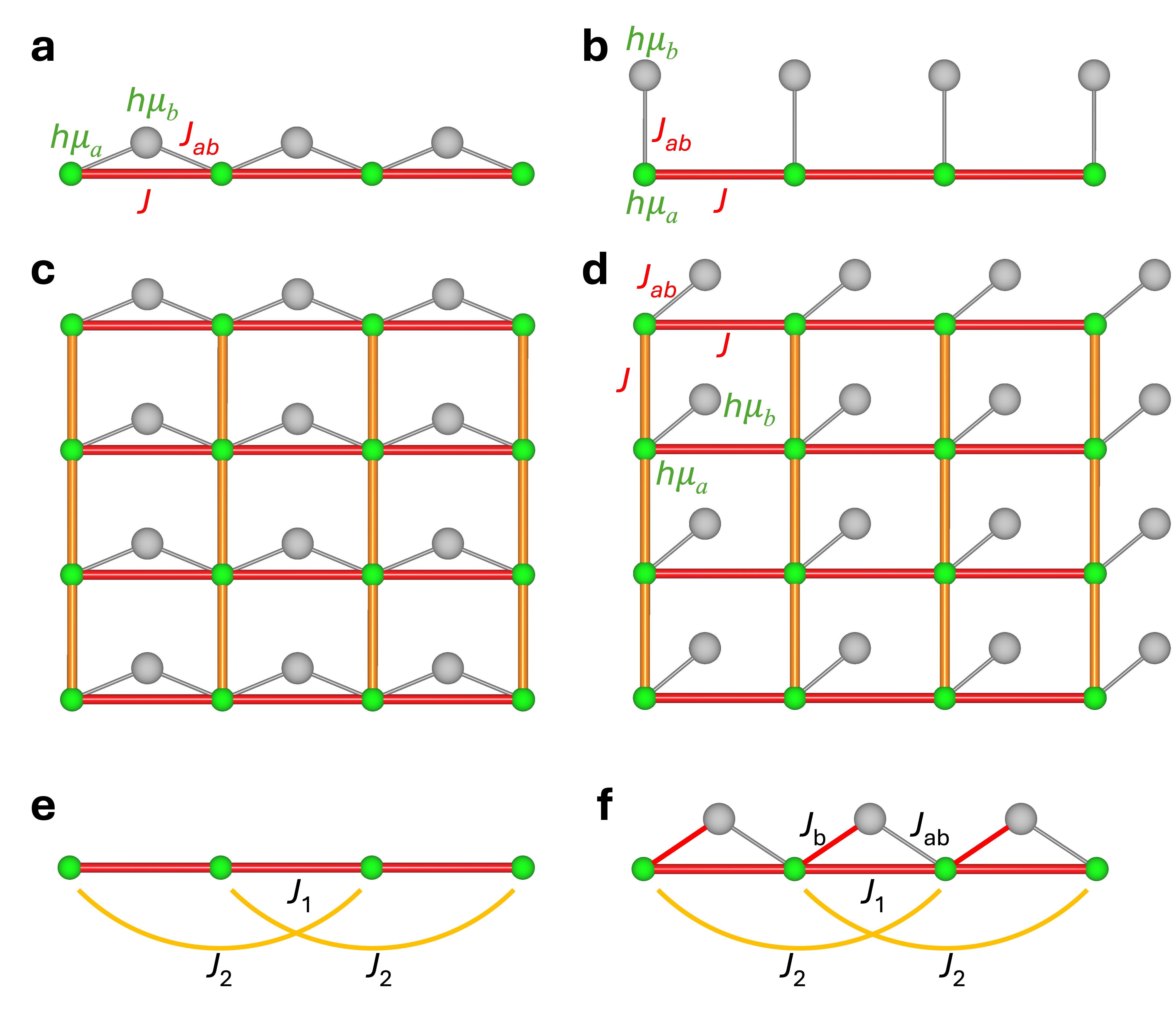}
    \end{center}
\caption{Schematics of minimally decorated Ising models in 1D and 2D: (a) and (c) bond decoration and (b) and (d) site decoration. Green balls depict the backbone, \typea spins coupled by the FM interaction $J$ (red and orange bonds); gray balls depict the decorating, \typeb spins coupled to the backbone by the AFM interaction $J_{ab}$ (gray bonds). The magnetic moments of the spins are $\mu_a$ and $\mu_b$, respectively. 
While the conventional geometric frustration is absent, \rred{frustration comes from the competition between an external field and the AFM coupling $J_{ab}$. (e) The zero-field $J_1$-$J_2$ Ising model. (f) The zero-field bond-decorated $J_1$-$J_2$ Ising model, where the decorating spins are coupled to two nearest-neighbor backbone spins by interactions $J_b$ and $J_{ab}$, yielding geometric frustration for $J_1 J_b J_{ab}<0$. Model (b) can be exactly mapped onto model (f) when $J_1=h\mu_a$, $J_b=h\mu_b$, and $J_2=J$.}
}
\label{Fig:structure}
\end{figure}

\red{Moreover, it was found that the in-field UNPC can occur in bond-decorated 1D Ising ferrimagnets without traditional geometric frustration (i.e., the frustration function has a positive sign)~\cite{Yin_MPT_chain}. The source of such unconventional frustration is the incompatibility between the magnetic field $h$ and antiferromagnetic (AFM) bonds, even in the zero-field limit $h\to 0$~\cite{Bell_JPC_74_Ising_ferri_1,Yin_g}. To unveil the underlying mechanism, we studied a minimal bond-decorated 1D Ising ferrimagnet without traditional geometric frustration [Fig.~\ref{Fig:structure}(a)], whose zero-temperature phase diagram features a half-fire, half-ice critical point where the backbone spins are fully disordered (fire) and the decorating spins are fully ordered (ice)~\cite{Yin_Ising_III_PRL,Yin_g}. This stimulated the idea of half-ice, half-fire as its twin state (where the backbone spins are fully ordered and the decorating spins are fully disordered), which however is an excited state everywhere in the ground-state phase diagram, i.e., it is hidden in the traditional approach. We looked into the excited-state phase diagram and identified that the half-ice, half-fire state emerges as a phase boundary between excited states with the needed macroscopic degeneracy that drives the UNPC: The disordering of the decorating spins drives the flipping of the fully ordered backbone spins. In this study, we also demonstrated that the use of a minimal bond-decorated model is instrumental to the understanding of the UNPC phenomenon thanks to the simplicity and extensibility of the minimal model to more complex models; for example, the UNPC in Ref.~\cite{Yin_MPT_chain} is now understood to be driven by a hidden 1/3-ice, 2/3-fire state.}

\red{Here, we further point out that since conventional geometric frustration is unnecessary for the ice-fire--driven UNPC, bond decoration itself is also unnecessary; without being limited by bond decoration, an even simpler minimal model may emerge.}
The purpose of this paper is to introduce site decoration as an alternative paradigm for exotic phases, phase transitions, and UNPC, where additional spins are placed on lattice sites rather than bonds, thereby completely eliminating conventional geometric frustration [Figs.~\ref{Fig:structure}(b) and \ref{Fig:structure}(d)]. \rred{Now frustration comes from the interesting local competition between an external field and the local AFM coupling.} \rrred{The site (vertex) decoration was briefly mentioned in a seminal study of the bond decoration and the star-triangle transformation of Ising models~\cite{Fisher_PR_59_Ising_transform}, yet to our knowledge, its role in spin frustration has not been explored.} Using the Ising model, we demonstrate exactly that the site-decorated ferrimagnet exhibits an half-ice, half-fire--driven UNPC at a fixed finite temperature in 1D\rrred{---while the crossover width narrowed exponentially as $J$ increases to facilitate the collective behavior}. Furthermore, while the 2D and three-dimensional (3D) site-decorated Ising models in an external magnetic field remain unsolved \rred{exactly}, we rigorously reveal an half-ice, half-fire--driven spin-reversal transition. This spin reversal can be induced by a slight change in temperature or \rred{the magnitude of} the external magnetic field \rred{(without changing its direction)}, even in the weak-field limit, suggesting a pathway for energy-efficient applications in data storage and processing. These findings establish site decoration as a viable framework for faster exploration of unique physical properties in unconventional frustrated magnets than bond decoration.
\blue{Moreover, the geometric compatibility of site and bond decorations significantly expands the potential for functional materials design.}

\red{In addition, the site-decorated model provides a rigorous testbed for artificial intelligence (AI) in science\rred{---notably, the Ising model and statistical physics themselves are fundamental building blocks underlying the 2024 Nobel Prize in Physics for the discoveries of the laureates that enable machine learning with artificial neural networks.} In this work, the analytic derivation of the present results was not only validated but also improved by \texttt{OpenAI o3-mini-high} reasoning model during the 1000-Scientist AI Jam Session (see Appendix~\hyperlink{AI}{A}), inspiring the further use of AI as scientific discoverer~\cite{Yin_Potts_J1-J2_1D,Yin_Potts_UNPC}.} \rred{Recently, \texttt{o3-mini-high} helped us find exact analytical solutions to a decades-old problem in statistical mechanics, namely the 1D $q$-state Potts model ($q=2$ corresponds to the Ising model) with nearest-neighbor ($J_1$) and next-nearest-neighbor ($J_2$) interactions [Fig.~\ref{Fig:structure}(e)]. Unexpectedly, by matching relevant mathematical subspaces, the 1D $J_1$-$J_2$ $q$-state Potts model exactly maps onto a simpler 1D $q$-state Potts model, with $J_2$ acting as the nearest-neighbor interaction and $J_1$ as an external field, for arbitrary $q=2,3,4,\dots,\infty$~\cite{Yin_Potts_J1-J2_1D}. The existence of infinitely many exact mappings raises the general question of to what extent the frustration induced by an external magnetic field corresponds to the geometrical frustration arising spontaneously from competing spin–spin interactions, echoing the OP similarities discussed above for spontaneous and in-field UNPCs. Indeed, we found that the present site-decorated Ising model in a magnetic field [Fig.~\ref{Fig:structure}(b)] can be exactly mapped onto a more complicated zero-field bond-decorated $J_1$–$J_2$ Ising model [Fig.~\ref{Fig:structure}(f)], thereby \rrred{unambiguously demonstrating the collective nature of the UNPC} and greatly facilitating our understanding of the unconventional frustration and physics in the site-decorated model.}

\rred{The rest of the article is organized as follows: In Sec.~\ref{Sec1D}, we present the minimal 1D site-decorated Ising model, its mapping onto a zero-field bond-decorated $J_1$-$J_2$ Ising model, and the exact solutions about its thermodynamic properties. In Sec.~\ref{Sec2D}, site decoration in higher dimensions is investigated. For a square lattice, we show exact results at the spin-reversal transition and \rrrred{the approximate temperature dependence of spin reversal in an exact-mapping Monte Carlo (MC) method~\cite{Strecka_PRB_23_decorated_2D}}. Section~\ref{SecDiscuss} addresses some immediate implications of the present paper on further theoretical and experimental research and development, including} material and device design, particularly in mixed $d$-$f$ compounds ~\cite{Kim_NM_22_review_ferri,Ramirez_25_SmMn2Ge2} \rrred{where isotropic Heisenberg spins are more common than Ising spins}. 

\section{One dimension\label{Sec1D}}

\subsection{Model}

The minimally site-decorated Ising model in 1D is shown in Fig.~\ref{Fig:structure}(b) and described by \begin{equation}
    H=H_a+ H_b, \label{minimal}
\end{equation}
where
\begin{subequations}
\label{HaHb}
\begin{eqnarray}
H_a&=&-J\sum_{i=1}^{N}\sigma_{i}\sigma_{i+1}-h\mu_a \sum_{i=1}^{N}\sigma_{i}, \label{ordinary} \\
H_b&=&-J_{ab}\sum_{i=1}^{N}\sigma_{i} b_i- h\mu_b \sum_{i=1}^{N}b_i.
\label{decor}
\end{eqnarray}
\end{subequations}
Here, $H_a$ describes the backbone of the single chain with $\sigma_{i}=\pm1$ (green balls, referred to as type-\emph{a} spins) and $J>0$ the FM interaction (red bonds). This is the model that was studied by Ising one century ago~\cite{Ising1925}. Also, $H_b$ describes the decorated parts, where $b_i=\pm1$ (gray balls, referred to as type-\emph{b} spins) is coupled to the $i$th type-\emph{a} spin with the AFM interaction $J_{ab}<0$ (gray bonds). Further, $h$ depicts the magnetic field and $\mu_a$ and $\mu_b$ the magnetic moments of type-\emph{a} and type-\emph{b} spins, respectively. The relationship of $\mu_b > \mu_a > 0$ is used to represent ferrimagnetism---the model is invariant under the sign changes that satisfy
$\mu_a\mu_b J_{ab} < 0$, e.g., $\mu_a=2$ and $\mu_b=-3$ in Sr$_3$CuIrO$_6$~\cite{Yin_PRL_Sr3CuIrO6,Yin_g}. Finally, $N$ is the total number of the unit cell with $\sigma_{N+1}\equiv\sigma_{1}$, $b_{N+1}\equiv b_{1}$, viz., periodic boundary condition. 

\subsection{Method}

We used the transfer matrix method~\cite{Kivelson_24_book_statistical,Mattis_book_08_SMMS,Kramers_Wannier_PR_41_transferMatrix} to exactly calculate the partition function $Z=\mathrm{Tr}\,\exp(-\beta H)$ and the free energy per unit cell $f=-\frac{1}{N\beta} \ln Z$ in the thermodynamic limit $N\to\infty$, where $\beta=1/(k_\mathrm{B}T)$ with $T$ being the absolute temperature and $k_\mathrm{B}$ the Boltzmann constant, and thermodynamic properties such as the entropy $S=-\partial f/\partial T$, the specific heat $C_v=T\partial S/\partial T$, the total magnetization $m=-\partial f/\partial h=\mu_a\langle\sigma_i\rangle+\mu_b\langle b_i\rangle$, and the magnetic susceptibility $\chi=\partial m/\partial h$. 

The site-decorating \typeb spins can be summed out exactly for any dimension~\rrred{\cite{Fisher_PR_59_Ising_transform}}, resulting in 
\begin{equation}
  Z=\exp(NA)\;\mathrm{Tr}\,\exp(-\beta H_\mathrm{eff}),\label{Z}    
\end{equation}
where 
\begin{equation}
    A=\frac{1}{2}\left(\ln \boxedp +\ln \boxedm\right),
    \label{A}
\end{equation}
with the shorthand notation 
\begin{equation}
    \boxed{\pm}=2\cosh(\beta h\mu_b \pm \beta J_{ab}).
    \label{eq:boxedpm}
\end{equation}
Here, $H_\mathrm{eff}$ is the effective Hamiltonian for the \typea spins
\begin{equation}
H_\mathrm{eff}=-J\sum_{i=1}^{N}\sigma_{i}\sigma_{i+1}-h_\mathrm{eff}\mu_a \sum_{i=1}^{N}\sigma_{i}, \label{Heff}    
\end{equation}
which is of the same form as the undecorated Ising model defined in Eq.~(\ref{ordinary})---with $h$ being replaced by a temperature-dependent effective magnetic field 
\begin{equation}
    h_\mathrm{eff}= h + \frac{1}{2\beta\mu_a}\left(\ln \boxedp- \ln \boxedm\right). \label{heff}
\end{equation}
Here, $h_\mathrm{eff}$ is independent of $J$. Unlike in the bond-decorated Ising model, $J$ is not renormalized in the site-decorated model, significantly simplifying the analysis. Equation~(\ref{Heff}) is ready to be solved~\cite{Kivelson_24_book_statistical,Mattis_book_08_SMMS}.

\subsection{\rred{Mapping}}

\rred{The above 1D site-decorated Ising model in an external magnetic field can be exactly mapped onto a 1D zero-field bond-decorated $J_1$-$J_2$ Ising model, where $J_1$ and $J_2$ are the nearest-neighbor and next-nearest-neighbor interactions between the backbone spins, respectively, and the decorating spins are coupled to two nearest-neighbor backbone spins by interactions $J_b$ and $J_{ab}$, when $J_1=h\mu_a$, $J_b=h\mu_b$, and $J_2=J$, as shown in Fig.~\ref{Fig:structure}(f).} 

\rred{\emph{Proof}: Consider the zero-field bond-decorated $J_1$-$J_2$ Ising model. The decorating spins can be summed out exactly for any dimension~\cite{Fisher_PR_59_Ising_transform,Syozi_Ising_2D}, resulting in the partition function 
\begin{equation}
\widetilde{Z}=\exp(N\widetilde{A})\;\mathrm{Tr}\,\exp(-\beta \widetilde{H}_\mathrm{eff}),    \label{Z2}
\end{equation} 
where 
\begin{equation}
    \widetilde{A}=\frac{1}{2}\left(\ln \widetilde{\boxedp} +\ln \widetilde{\boxedm}\right),
    \label{A2}
\end{equation}
with the shorthand notation $\widetilde{\boxed{\pm}}=2\cosh(\beta J_b \pm \beta J_{ab} )$. Here, $\widetilde{H}_\mathrm{eff}$ is the effective Hamiltonian for the backbone spins
\begin{equation}
\widetilde{H}_\mathrm{eff}=-J_1^\mathrm{eff}\sum_{i=1}^{N}\sigma_{i}\sigma_{i+1}-J_2 \sum_{i=1}^{N}\sigma_{i}\sigma_{i+2}, \label{Heff2}    
\end{equation}
which is of the same form as the standard $J_1$-$J_2$ Ising model [Fig.~\ref{Fig:structure}(e)]---with $J_1$ being replaced by a temperature-dependent effective nearest-neighbor interaction 
\begin{equation}
    J_1^\mathrm{eff}= J_1 + \frac{1}{2\beta}\left(\ln \widetilde{\boxedp}- \ln \widetilde{\boxedm}\right). \label{Jeff}
\end{equation}
}

\rred{Next, with the substitution of $\tau_i=\sigma_i\sigma_{i+1}$ and $\tau_i\tau_{i+1}=\sigma_i\sigma_{i+2}$, the 1D $J_1^\mathrm{eff}$-$J_2$ Ising model can be exactly mapped onto the 1D Ising model with $J_2$ acting as the nearest-neighbor interaction and $J_1^\mathrm{eff}$ as an external field~\cite{Yin_Potts_J1-J2_1D,Dobson_JMathP_69_Many-Neighbored-Ising-Chain}, i.e.,
\begin{equation}
\widetilde{H}_\mathrm{eff}=-J_2\sum_{i=1}^{N}\tau_{i}\tau_{i+1}-J_1^\mathrm{eff} \sum_{i=1}^{N}\tau_{i}, \label{Heff3}    
\end{equation}
where $\tau_i=\pm1$ is the Ising spin values. 
Therefore, Eq.~(\ref{Heff3}) can be exactly mapped onto Eq.~(\ref{Heff}) when $J_2=J$, $J_1^\mathrm{eff}=h_\mathrm{eff}\mu_a$, and $J_b=h\mu_b$, leading to $\widetilde{A}=A$ and $J_1=h\mu_a$. This also means $\widetilde{Z}=Z$. Hence, the 1D zero-field bond-decorated $J_1$-$J_2$ Ising model has been mapped exactly onto the 1D site-decorated Ising model in an external magnetic field. \qedblack}

\rred{The bond-decorated $J_1$–$J_2$ model exhibits conventional geometrical frustration when $J_1 J_b J_{ab}<0$, i.e., when the frustration function for any triangle formed by the three interactions is negative~\cite{toulouse1977frustration}. Correspondingly, the site-decorated model is frustrated for $\mu_a\mu_b J_{ab}<0$. In addition, for a UNPC to occur, $J_1$ must be the weakest among all interactions, particularly when $|J_1|\ll |J_2|$ (see the next section for proof). Such a condition is not easily realized for spin–spin interactions. By contrast, in the site-decorated model it is straightforward to achieve $|h\mu_a|\ll |J|$ with a weak magnetic field $h$. Therefore, the mapping offers the dual advantage of employing conventional geometrical frustration to elucidate the origin of frustration and collective behavior, while using the magnetic field as a convenient knob for tuning the parameters.}

\subsection{Results}

\begin{figure}[t]
    \begin{center}
\includegraphics[width=0.99\columnwidth,clip=true,angle=0]{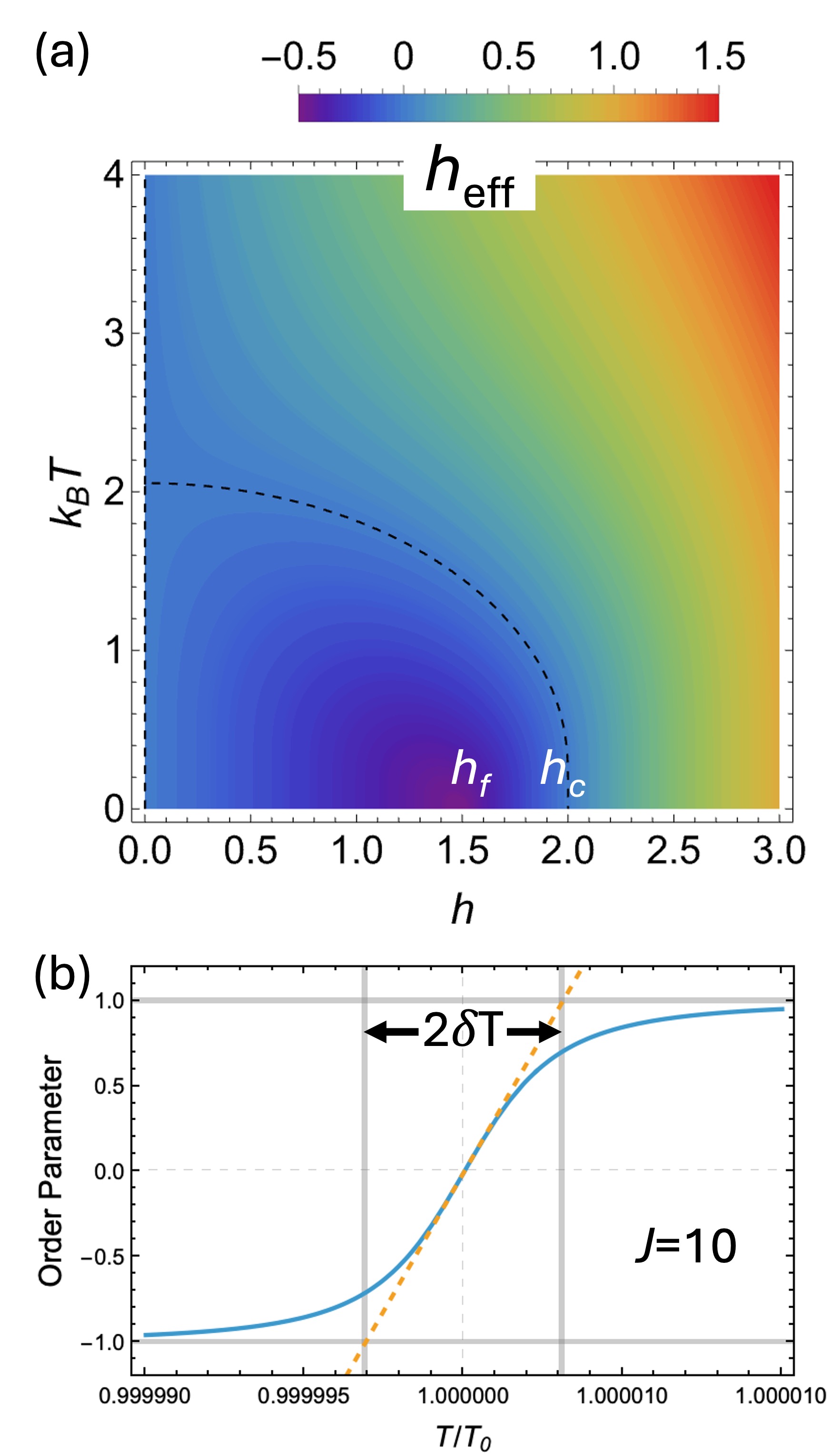}
    \end{center}
    \vspace{-0.4cm}
\caption{(a) Density plot of  $h_\mathrm{eff}$ in the $h$-$T$ plane. The \rred{dashed} line is the $h_\mathrm{eff}=0$ contour line, i.e., the $T_0$ curve given by Eq.~(\ref{eq:T0}). \rred{(b) OP $\langle \sigma_i\rangle$ and the definition of the crossover width $2\delta T = 2\left(\partial \langle \sigma_i \rangle/\partial T\right)^{-1}_{T=T_0}$. Here, $J_{ab}=-2$, $\mu_a=1$, and $\mu_b=4/3$, resulting in $h_c\equiv|J_{ab}/\mu_a|=2$ and $h_f\equiv|J_{ab}/\mu_b|=1.5$.}}
\label{Fig:heff}
\vspace{-0.2cm}
\end{figure}


\begin{figure*}[t]
    \begin{center}
\includegraphics[width=0.97\textwidth,clip=true,angle=0]{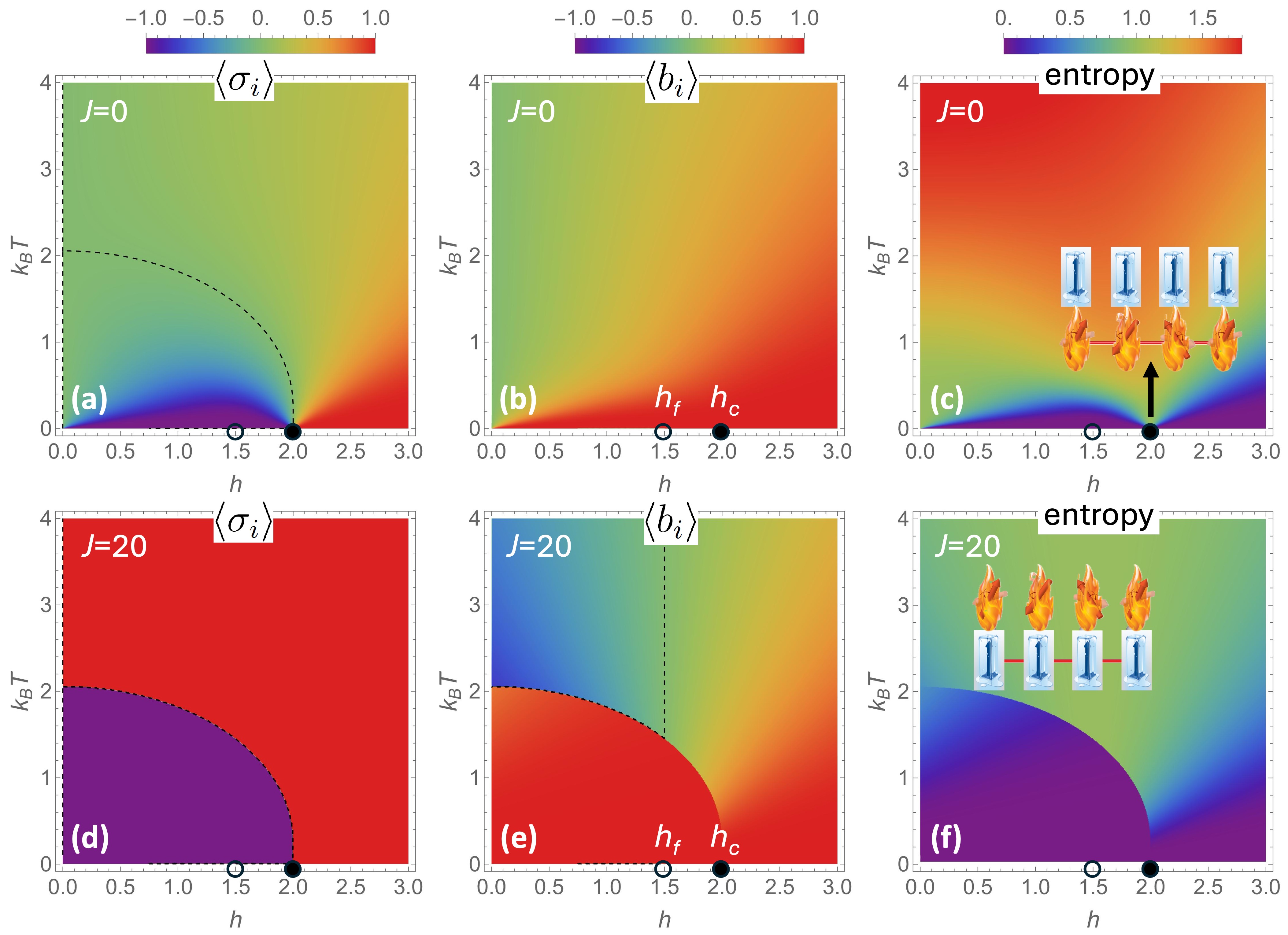}
   \end{center}
\caption{\rred{The density plots of $\langle\sigma_i\rangle$, $\langle b_i\rangle$, and entropy $S/\ln2$ in the $h$-$T$ plane. Top panels: (a)--(c) for $J=0$. Bottom panels: (d)--(f) for $J=20$. Here $J_{ab}=-2$, $\mu_a=1$, and $\mu_b=4/3$, resulting in $h_c=2$ (solid circle) and $h_f=1.5$ (open circle). The dashed lines in (a),(d),(e) denote the contour of zero value: In (a), the $\langle\sigma_i\rangle=0$ line defines $T_0$ the phase boundary in (d) by Eq.~(\ref{eq:T0}). In (c), the black arrow points to the zero-temperature critical point at $h=h_c$ hosting the half-fire, half-ice state. In (e), the vertical line $\langle b_i\rangle=0$ at $h=h_f$ and above the phase boundary line hosts the opposite half-ice, half-fire state---shown in (f)---originates from the hidden frustration, and divides the region above $T$=$T_0$ in half where the decorated spins flip for $h<h_f$.}}
\label{Fig:phasediagram}
\end{figure*}

\subsubsection{Ultranarrow phase crossover}

\begin{figure*}[th]
    \begin{center}
\includegraphics[width=0.9\textwidth,clip=true,angle=0]{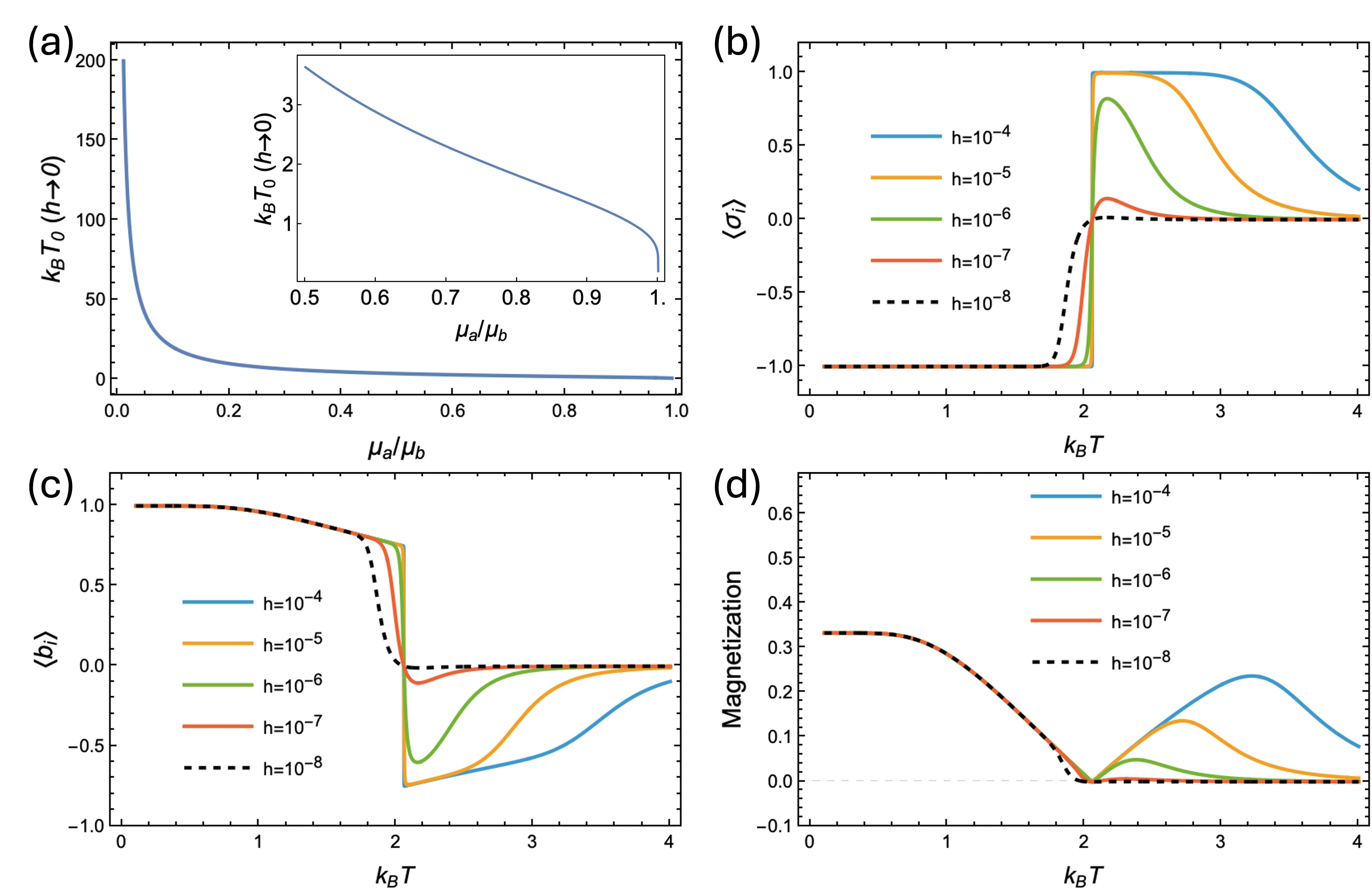}
    \end{center}
\caption{Approaching the zero field. (a) $T_0$ as a function of $\mu_a/\mu_b$ for $h\to0$. The temperature dependence of (b) $\langle\sigma_i\rangle$, (c) $\langle b_i\rangle$, and (d) the total magnetization $m=\mu_a\langle\sigma_i\rangle+\mu_b\langle b_i\rangle$ in weak fields. $J=20$, $J_{ab}=-2$, $\mu_a=1$, and $\mu_b=4/3$ unless specified.}
\label{Fig:approach_h0}
\end{figure*}

The OP that allows for fast and accurate identification of UNPC is the magnetization of the backbone spins given by $\langle \sigma_i \rangle=-\frac{\partial f}{h\,\partial \mu_a}
$~\cite{Yin_MPT_chain,Yin_Ising_III_PRL}
\begin{equation}
    \langle \sigma_i \rangle
    =\frac{\sinh(\beta  h_\mathrm{eff}\mu_a)}{\sqrt{\sinh^2(\beta h_\mathrm{eff}\mu_a)+\exp(-4\beta J)}}, 
    \label{eq:sigma}
\end{equation} 
which switches sign at the crossover temperature $T_0$. Then according to Eq.~(\ref{eq:sigma}), $h_\mathrm{eff}$ must switch sign at $T_0$. This means that $T_0$ can be determined by the following condition:
\begin{equation}
    h_\mathrm{eff}=0 \;\;\mathrm{at}\;\; T_0,
\end{equation}
yielding the critical equation
\red{
\begin{equation}
\tanh(-\beta_0 J_{ab}) =\frac{\tanh(\beta_0 h\mu_a)}{\tanh(\beta_0 h\mu_b)},
\label{eq:T0}
\end{equation}
where $\beta_0\equiv1/(k_\mathrm{B}T_0)$.
This form of the critical equation was derived by the OpenAI reasoning model \texttt{o3-mini-high}  (see Appendix~\hyperlink{AI}{A}), which elegantly links four parameters of the Hamiltonian but is independent of $J$. Equation~(\ref{eq:T0}) has a finite-temperature solution for $|\mu_a| < |\mu_b|$,  $\mu_a\mu_bJ_{ab}<0$, and $0<h<h_c$ or $-h_c<h<0$, where $h_c\equiv|J_{ab}/\mu_a|$, as shown in Fig.~\ref{Fig:heff}(a)} \rred{where the dashed contour line means $h_\mathrm{eff}=0$. These constraints on the parameters can be understood as $J_1J_bJ_{ab}<0$, $|J_1|<|J_b|$, and $|J_1|<|J_{ab}|$ in the equivalent 1D zero-field bond-decorated $J_1$-$J_2$ Ising model, i.e., $J_1$ is the weakest among the three interactions that form a geometrically frustrated triangle [Fig.~\ref{Fig:structure}(f)].}

\rred{The result that $T_0$ is independent of $J$ means that the same $T_0$ curve holds for $J=0$, as shown in Fig.~\ref{Fig:phasediagram}(a). Since for $J=0$ the system is decoupled into pairs of one type-$a$ spin and one type-$b$ spin, it is clear that the frustration in the site-decorated model comes from the interesting local competition between an external field and the local AFM coupling. However, there is no sharp phase switch crossing the $T_0$ curve in Fig.~\ref{Fig:phasediagram}(a). Thus, in addition to finite $T_0$, the realization of an UNPC also requires a mechanism to quickly narrow the crossover width $2\delta T$, which is defined as
~\cite{Yin_MPT_chain,Yin_Ising_III_PRL}
\begin{eqnarray}
    \delta T&=&\left(\frac{\partial \langle\sigma_i\rangle}{\partial T}\right)^{-1}_{T=T_0}
\end{eqnarray}
and illustrated in Fig.~\ref{Fig:heff}(b). One finds
\begin{eqnarray}
    \delta T&=&\chi_0^{-1}\left(\frac{\partial h_\mathrm{eff}}{\partial T}\right)^{-1}_{T=T_0} \nonumber\\
    &=& \chi^{-1}_0 \left(-T_0\mu_a\right) \nonumber \\
    && \times\frac{1}{2}[2h\mu_a  
    +(h\mu_b+J_{ab})\tanh(\beta_0 h\mu_b + \beta_0 J_{ab}) \nonumber \\
    &&-(h\mu_b-J_{ab})\tanh(\beta_0 h\mu_b - \beta_0 J_{ab})]^{-1},
    \label{eq:dT}
\end{eqnarray}
where $\chi_0$ is the initial magnetic susceptibility of the effective Hamiltonian [Eq.~(\ref{Heff})] at $T_0$ given by
\begin{eqnarray}
    \chi_0=\left.\frac{\partial \langle\sigma_i\rangle}{\partial h_\mathrm{eff}}\right|_{h_\mathrm{eff}=0}=\exp(2\beta_0J)\beta_0\mu_a.
\end{eqnarray}
Hence,} increasing $J$ will not affect $T_0$ but can exponentially narrow the crossover by $2\delta T \propto \exp(-2\beta_0 J)$ \rred{thanks to the large value of $\chi_0$}.

As shown in Figs.~\ref{Fig:phasediagram}(d)--\ref{Fig:phasediagram}(f), the essential physics of UNPC---driven by the hidden half-ice (\typea spins are fully ordered), half-fire (\typeb spins are fully disordered) state in the bond-decorated Ising model~\cite{Yin_Ising_III_PRL}---has been reproduced in the site-decorated model. In this framework, $\langle \sigma_i \rangle$ flips from $+1$ to $-1$ at a fixed $T_0$, while $2\delta T$ narrows exponentially as $J$ increases.

\blue{To gain more insight, the magnetization of the decorating spins is given by $\langle b_i \rangle = -\frac{\partial f}{h\,\partial \mu_b}$
\begin{eqnarray}
    \langle b_i \rangle = &&\frac{1+\langle \sigma_i \rangle}{2}\tanh[\beta \mu_b (h-h_f)] \nonumber\\ &&
    +\frac{1-\langle \sigma_i \rangle}{2}\tanh[\beta \mu_b (h+h_f)],
    \label{eq:mb}
\end{eqnarray}
where $h_f\equiv -J_{ab}/\mu_b$. For sufficiently large $J>0$, outside the ultranarrow crossover region, 
\begin{equation}
    \langle b_i \rangle  \approx\left\{
\begin{array}{lcc}
  \tanh[\beta\mu_b(h + h_f)] > 0\; &\mathrm{for}& \; T < T_0 \\
  \tanh[\beta\mu_b (h-h_f)]\; &\mathrm{for}& \;\;\; T > T_0 \\
\end{array}
\right. . \label{eq:OP_b}
\end{equation}
For $T>T_0$, $\langle b_i \rangle  \approx 0$ at $h=h_f$, which is the phase boundary between two excited states in the zero-temperature phase diagram, indicates that the \typeb spins are disordered (on fire). 
}

\blue{For completeness, $ \langle b_i \rangle$ for the bond-decorated model is given by 
\begin{eqnarray}
    \langle b_i \rangle = &&
    \frac{1+2\langle \sigma_i \rangle+\langle \sigma_i \sigma_j\rangle}{4}\tanh[\beta \mu_b (h-h^\mathrm{bond}_f)]  \nonumber\\ &&
   + \frac{1-2\langle \sigma_i \rangle+\langle \sigma_i \sigma_j\rangle}{4}\tanh[\beta \mu_b (h+h^\mathrm{bond}_f)]  \nonumber\\ &&
   + \frac{1-\langle \sigma_i \sigma_j\rangle}{2}\tanh(\beta\mu_b h),
    \label{eq:mb_bond}
\end{eqnarray}
with $h^\mathrm{bond}_f\equiv -2J_{ab}/\mu_b$ since each \typeb spin is coupled to two \typea spins in bond decoration. For sufficiently large FM $J$ that ensures $\langle \sigma_i \sigma_j\rangle\simeq1$, Eq.~(\ref{eq:mb_bond}) reduces to Eq.~(\ref{eq:mb}) with $h_f\equiv -J_{ab}/\mu_b$ in the latter being replaced by $h^\mathrm{bond}_f$.} 
\rred{The critical equation for the bond-decorated model is determined by
\begin{equation}
\tanh(-2\beta J_{ab}) =\frac{\tanh(\beta h\mu_a)}{\tanh(\beta h\mu_b)}.
\label{eq:T0_bond}
\end{equation}
Therefore, the results of the site-decorated model with $J_{ab}=-2$ can reproduce the results of the bond-decorated model with $J_{ab}=-1$ for large $J$. 
}

\rred{The degree of frustration can be quantified by large entropy contributed from the on-fire spins. For $J=0$, the frustration is represented by the backbone spins, which are on fire at the zero-temperature critical point $h=h_c$ while the decorating spins are frozen, as shown by the half-fire, half-ice state in Fig.~\ref{Fig:phasediagram}(c). By contrast,  increasing $J$ will quickly transfer the frustration to the decorating spins, which are on fire at $T_0$, while the backbone spins are frozen, resulting in a half-ice, half-fire state with the entropy per unit cell $S=\ln 2$, as shown in Fig.~\ref{Fig:phasediagram}(f).}

\subsubsection{\rred{The weak $h$ limit}}

As $h$ increases, both $T_0$ and $\delta T$ decrease, reaching $T_0=0$ and $2\delta T=0$ at $h_c$. The maximum $T_0$ as a function of $h$ occurs at the weak field limit $h\to 0$, 
\begin{equation}
k_\mathrm{B}T^\mathrm{max}_0=2J_{ab}\left(\ln\frac{1 - \mu_a/\mu_b}{1 + \mu_a/\mu_b}\right)^{-1}. \label{eq:T0max}
\end{equation}
As shown in Fig.~\ref{Fig:approach_h0}(a), $T^\mathrm{max}_0$ rapidly increases as $\mu_a/\mu_b$ decreases, a feature desirable for low-field, high-temperature applications. On the other hand, as shown in the inset of Fig.~\ref{Fig:approach_h0}(a), $T^\mathrm{max}_0$ changes mildly for a large range of $\mu_a/\mu_b$. This is good news for the $\exp(-2\beta_0 J)$ factor that narrows $\delta T$.

Note that at the absolute zero field ($h=0$), there is no sign change in $h_\mathrm{eff}$ because $h_\mathrm{eff}=0$ and $\langle \sigma_i \rangle=0$ hold for any finite temperature, \rred{as shown by the vertical dashed lines at $h=0$ in Figs.~\ref{Fig:heff}(a), \ref{Fig:phasediagram}(a), and \ref{Fig:phasediagram}(d). Hence, $T_0$ is ill defined at $h=0$}. This is consistent with the theorem that spontaneous UNPC does not exist in single-chain Ising models due to the spin-reversal symmetry associated with $h=0$~\cite{Yin_MPT}. The minimal model for spontaneous UNPC is a decorated two-leg ladder Ising model~\cite{Yin_MPT,Yin_icecreamcone}. 

\rred{To understand the abrupt change in $T_0$ between $h=0$ and $h\to 0$, we derive $h_\mathrm{eff}$ and
the crossover width in the weak $h$ limit:
\begin{eqnarray}
\lim_{h\to 0} h_\mathrm{eff} &=& h\left[1+\frac{\mu_b}{\mu_a}\tanh(\beta J_{ab})\right],\label{eq:heffh0}\\
\lim_{h\to 0} \frac{\delta T}{T_0} &=&\frac{\exp(-2\beta_0 J)}{\beta_0 h\mu_a}\frac{\sinh(2\beta_0 J_{ab})}{2\beta_0J_{ab}}.
\label{eq:dTh0}
\end{eqnarray}
Here, $\delta T$ diverges as $h\to 0$ for fixed $J$. This means the absence of UNPC even with a finite $T_0^\mathrm{max}$, in agreement with the aforementioned theorem.}

\rred{For $h \to 0$, the crossover width grows with $1/h$ in the form of $\delta T(h,J)=\delta T(1,0)\exp(-2\beta_0 J)/h$. As shown in Figs.~\ref{Fig:approach_h0}(b)--\ref{Fig:approach_h0}(d), the substantial changes in  $\langle\sigma_i\rangle$, $\langle b_i\rangle$, and the total magnetization $m=\mu_a\langle\sigma_i\rangle+\mu_b\langle b_i\rangle$ still exist at $T_0$---a V-shaped $m$ with the minimum at $T_0$ is interesting---until lost for sufficiently weak $h$. Suppose $2\Delta T$ is the targeted crossover width such as the resolution of a measuring instrument, i.e., $2\delta T(h,J) \le 2\Delta T$, we get $h\ge \delta T(1,0)\exp(-2\beta_0 J)/\Delta T$, i.e., there exists a threshold field $h_0(J)\equiv\delta T(1,0)\exp(-2\beta_0 J)/\Delta T$, below which the UNPC vanishes. It is also clear that $J$ can make the threshold exponentially negligible.
}

However, the limitation due to the divergence of $2\delta T$ as $h\to 0$ can be overcome in higher dimensions where genuine finite-temperature phase transitions are known to exist at $h=0$, as demonstrated below. Then the different responses of the system to $h=0$ and $h\to 0$ can be used to measure the absolute zero magnetic field.

\ignore{
\begin{figure}[t]
    \begin{center}
\includegraphics[width=0.97\columnwidth,clip=true,angle=0]{fig4_M2D.jpeg}
    \end{center}
\caption{\label{fig:spin-reversal}The reversal of $\langle \sigma_i\rangle$ at $T_0$ as $h_\mathrm{eff}\to 0^\pm$ in the site-decorated square-lattice Ising model for $J=1$ with $T_0^\mathrm{max}<T_c^*\approx 2.269/k_\mathrm{B}$ (blue solid and dashed lines) and $J=0.88$ with $T_0^\mathrm{max} > T_c^*\approx 2/k_\mathrm{B}$ (red solid and dashed lines). Here $J_{ab}=-2$, $\mu_a=1$, $\mu_b=4/3$ yielding $T_0^\mathrm{max}\approx 2.056/k_\mathrm{B}$.}
\label{Fig:2D}
\end{figure}
}

\begin{figure}[t]
    \begin{center}
    \ignore{
        \subfigure[]{
\includegraphics[width=0.97\columnwidth,clip=true,angle=0]{fig4_M2D.jpeg}
        }
        \subfigure[]{
\includegraphics[width=0.97\columnwidth,clip=true,angle=0]{fig6a_Mb2D.jpeg}
        }
        \subfigure[]{
\includegraphics[width=0.97\columnwidth,clip=true,angle=0]{fig6b_Mtot2D.jpeg}
        }
        }
\includegraphics[width=0.95\columnwidth,clip=true,angle=0]{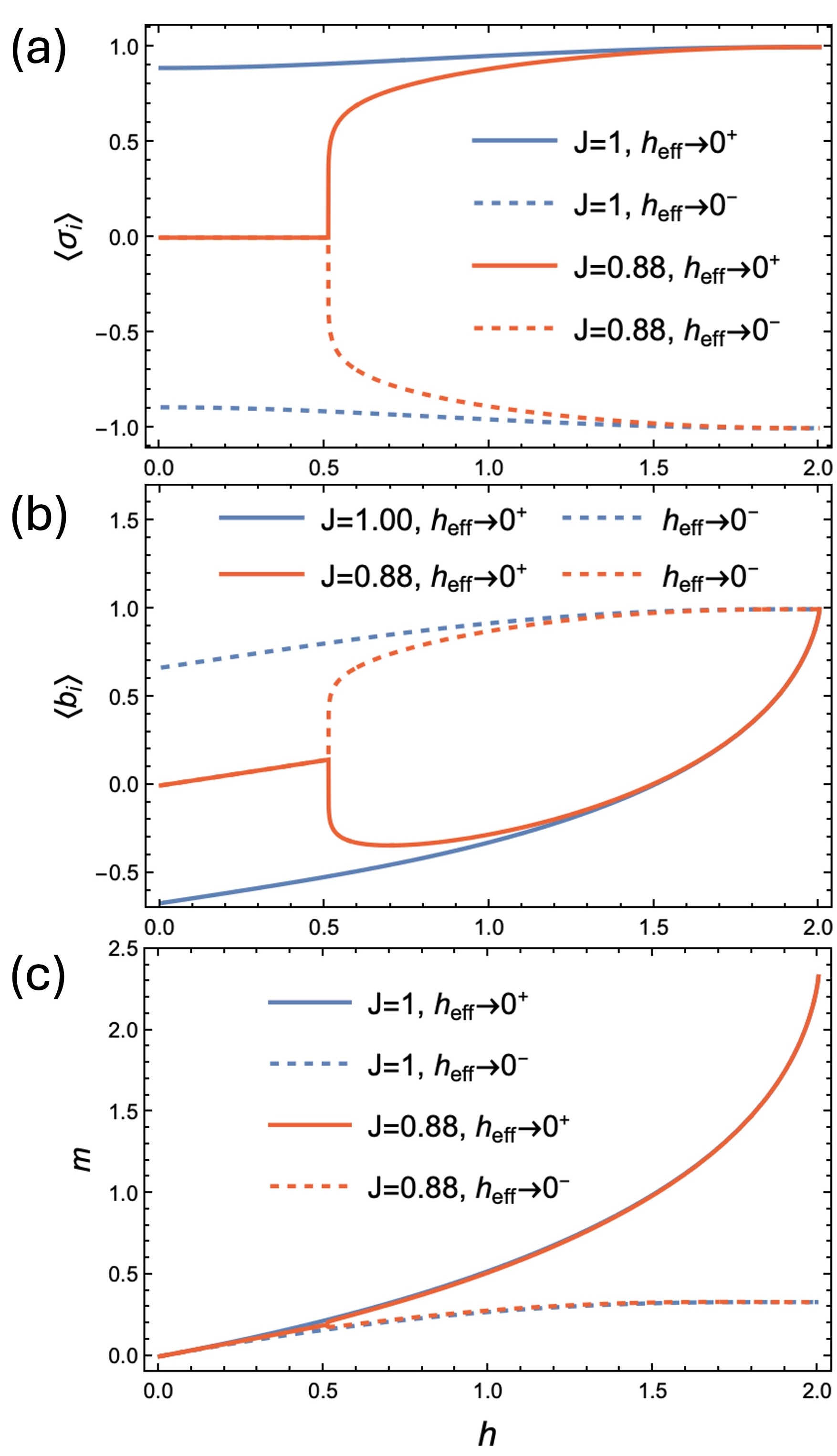}    
\end{center}
\caption{The magnetic field dependence of (a) $\langle \sigma_i\rangle$, \blue{(b) $\langle b_i \rangle$, and (c) the total magnetization $m=\mu_a\langle \sigma_i \rangle + \mu_b\langle b_i \rangle$} at $T_0$ as $h_\mathrm{eff}\to 0^\pm$ in the site-decorated square-lattice Ising model for $J=1$ with $T_c^*\approx 2.269/k_\mathrm{B} > T_0^\mathrm{max}$ (blue solid and dashed lines) and $J=0.88$ with $ T_c^*\approx 2/k_\mathrm{B} < T_0^\mathrm{max}$ (red solid and dashed lines). Here $J_{ab}=-2$, $\mu_a=1$, $\mu_b=4/3$ yielding $T_0^\mathrm{max}\approx 2.056/k_\mathrm{B}$.}
\label{Fig:2D} 
\end{figure}

\section{Higher dimensions\label{Sec2D}}

\subsection{Exact results}

Although 2D and 3D Ising models with a nonzero magnetic field remain unsolved \rred{exactly}, it has been proven for any dimension that site decoration yields $h_\mathrm{eff}=0$ at $T_0$ as a function of $h$ [the black line in Fig.~\ref{Fig:phasediagram}(a)], while keeping $J$ unchanged. Equations~(\ref{A})--(\ref{heff}), (\ref{eq:T0}), (\ref{eq:mb}), and (\ref{eq:T0max}) are directly applicable to the 2D and 3D cases. Consequently, the following exact result can be derived as a function of $h$.

\ignore{
The exact free energy per unit cell of the rectangular lattice in the thermodynamic limit for $h_\mathrm{eff}=0$ is given by~\cite{Onsager_Ising_2D,Mattis_book_08_SMMS}
\begin{eqnarray}
    f&=&-k_\mathrm{B}T\Big\{A+\ln2+\frac{2}{(2\pi)^2}\int_{0}^{\pi}d\theta \int_{0}^{\pi}d\phi\nonumber\\
    &\,&\ln[\cosh2K_1\cosh2K_2 \nonumber\\ 
    &\;\;\;\;\;\;&-\sinh2K_1\cos\theta 
    -\sinh2K_2\cos\phi]\Big\},
    \label{Onsager:free}
\end{eqnarray}
where $K_1=\beta J_1$ and $K_2=\beta J_2$ for shorthand notation. The paramagnetic-ferromagnetic phase transition takes place at the logarithmic divergence in Eq.~(\ref{Onsager:free}) when 
\begin{equation}
    \cosh2K_1\cosh2K_2 
    -\sinh2|K_1|-\sinh2|K_2|=0,
    \label{Onsager:Tc}
\end{equation}
which determines the transition temperature $T_c^*$ and can be simplified to 
\begin{equation}
    \sinh2|K_1| \sinh2|K_2|=1, 
    \label{Onsager:Tc2}
\end{equation}
or more easily to be visualized
\begin{equation}
   e^{-2|K_1|}=\tanh|K_2|. 
   \label{Onsager:Tc3}
\end{equation} 

For a square lattice with $K_1=K_2=\beta J$~\cite{Kramers_Wannier_PR_41_transferMatrix},}

For a site-decorated square lattice [Fig.~1(d)], the system reduces to a simple square lattice with $h_\mathrm{eff}=0$ at $T_0$, which would undergo a spontaneous paramagnetic-FM phase transition at~\cite{Kramers_Wannier_PR_41_transferMatrix,Onsager_Ising_2D}
\begin{equation}
k_\mathrm{B}T_c^*=2J/\ln{(1+\sqrt{2})}\approx 2.2692J,
\label{eq:Tc}
\end{equation} 
with the OP for $h_\mathrm{eff}\to 0$ given by~\cite{Yang_Ising_2D_M,Mattis_book_08_SMMS}
\begin{equation}
 \langle \sigma_i \rangle = \mathrm{sgn}(h_\mathrm{eff})\left[1 - (\sinh2 \beta_0J)^{-4}\right]^{1/8},
\label{Onsager:m2}
\end{equation}
for $T_0<T_c^*$ and zero for $T_0 > T_c^*$, where $\beta_0=1/(k_\mathrm{B}T_0)$. 

Hence, the sign change of $h_\mathrm{eff}$ at $T_0$ induces a spin reversal transition. Recall that this sign change can be triggered by a slight variation in temperature or magnetic field along the $T_0$ curve [the black line in Fig.~\ref{Fig:phasediagram}(a)]. As shown in Fig.~\ref{Fig:2D}(a), two cases arise: (i) If $T_0^\mathrm{max}<T_c^*$ (blue solid and dashed lines), spin reversal occurs throughout the entire region $0<h<h_c$. (ii) If $T_0^\mathrm{max}>T_c^*$ (red solid and dashed lines), spin reversal has a threshold in $h$, where the corresponding $T_0$ equals $T_c^*$; \red{the threshold $h$ always exist because $T_0$ continuously decreases to zero as $h$ increases to $h_c$. These two cases imply that the 2D/3D spin reversal transition puts no constraint on $J$, unlike UNPC in 1D, which requires small $\exp(-2J/k_\mathrm{B}T_0)$.}

\blue{The magnetic field dependence of $\langle b_i \rangle$ at $T_0$ as $h_\mathrm{eff}\to 0^\pm$ is plotted in Fig.~\ref{Fig:2D}(b). It qualitatively follows Eq.~(\ref{eq:OP_b}); $\langle b_i \rangle  \approx 0$ at $h=h_f$ for $h_\mathrm{eff}\to 0^+$ indicates that the \typeb spins are disordered (on fire).}

\blue{The magnetic field dependence of the total magnetization per unit cell $m=\mu_a\langle \sigma_i \rangle + \mu_b\langle b_i \rangle$ at $T_0$ as $h_\mathrm{eff}\to 0^\pm$ is plotted in Fig.~\ref{Fig:2D}(c). The small difference as $h_\mathrm{eff}$ changes from $0^-$ to $0^+$ for $h<0.5$ suggests that the weak-field-driven reversal of the \typea spins is energy efficient. On the other hand, $m$ changes a lot as $h_\mathrm{eff}$ changes from $0^-$ to $0^+$ for $1<h<2$, implying a large magnetocaloric effect and its potential application for magnetic refrigeration.} 



Similarly, the spin reversal transition at 
$h_\mathrm{eff}=0$ also occurs in other site-decorated 2D and 3D Ising models under an external magnetic field. This transition follows two cases analogous to Fig.~\ref{Fig:2D}, with exact results calculable for the anisotropic rectangular, anisotropic triangular~\cite{Potts_PR_52_Ising_triangular_magnetization}, honeycomb, and kagom\'{e}~\cite{Naya_54_Ising_honeycomb_kagome_magnetization} lattices, where the exact expression for spontaneous magnetization---akin to Eq.~(\ref{Onsager:m2}) for a square lattice---is known.

\rrrred{\subsection{Exact-mapping MC method}\label{hybrid}}

\rrrred{To illustrate the temperature dependence of spin reversal in the 2D site-decorated Ising model, we adopt an exact-mapping MC method~\cite{Strecka_PRB_23_decorated_2D} in which the decorating spins are treated exactly---using Eqs.~(\ref{A})--(\ref{heff}), (\ref{eq:T0}), (\ref{eq:mb}), and (\ref{eq:T0max})---and the backbone spins are accurately treated in the MC method with the exact effective field.}

\begin{figure}[t]
    \begin{center}
\includegraphics[width=\columnwidth,clip=true,angle=0]{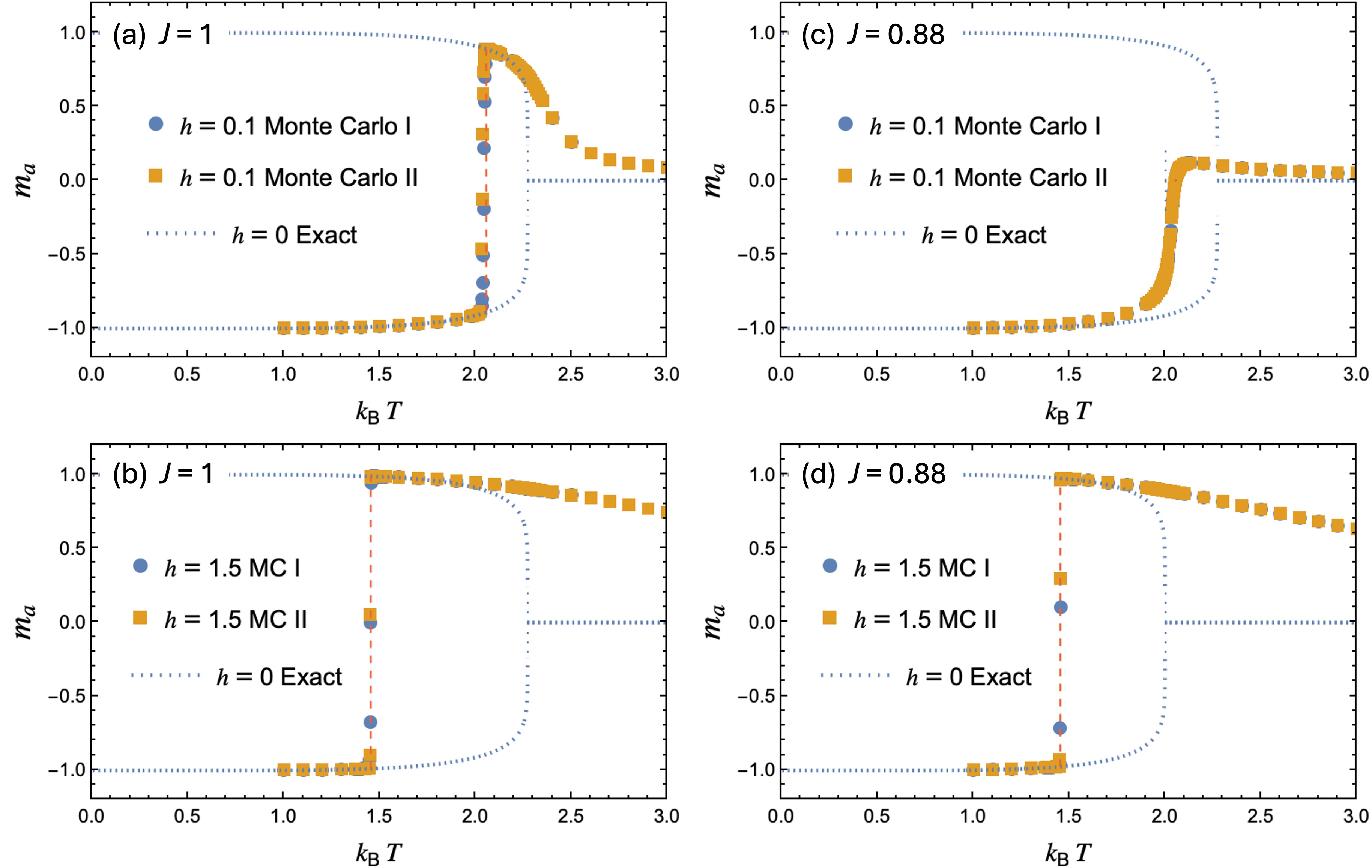}
    \end{center}
\caption{\rrrred{The $T$ dependence of $m_a(h,T)$ for $h=0.1$ and $1.5$ in the exact-mapping MC method for the site-decorated Ising model on a square lattice with linear size $L=120$ for (a) and (b) $J=1$, where $T_0^\mathrm{max}<T_c^*$, and (c) and (d) $J=0.88$, where $T_0^\mathrm{max}>T_c^*$. The other model parameters are the same as those in Fig.~\ref{Fig:2D}. Blue dots are MC results for $n_\mathrm{therm}=10\,000$ and $n_\mathrm{measure}=50\,000$. Orange dots are MC results for $n_\mathrm{therm}=20\,000$ and $n_\mathrm{measure}=100\,000$. Blue dotted lines are Onsager's exact results for $h=0$. Red dashed lines mark the spin reversal transition at $T_0(h)$, as shown exactly in Fig.~\ref{Fig:2D}.}}
\label{Fig:2D_MC}
\end{figure}

\rrrred{We simulate the effective square-lattice Ising backbone---with linear size $L=120$~\cite{Strecka_PRB_23_decorated_2D} and periodic boundary conditions---obtained after tracing out the decorated spins, which yields a temperature-dependent effective field $h_{\rm eff}(h,T)$ [Eq.~(\ref{heff}]. For each bare field $h$, we first determine the crossover temperature $T_0(h)$ as the unique root of $h_{\rm eff}(h,T_0)=0$ [Eq.~(\ref{eq:T0})], using a bracketing/bisection solver. We then construct a nonuniform temperature grid $T\in[T_{\min},T_{\max}]$ that is refined around both $T_0(h)$ and $T_c^*$ [Eq.~(\ref{eq:Tc})] and run a parallel-tempering (replica-exchange) simulation in temperature. Each replica at temperature $T_r$ evolves with a single-spin heat-bath (Glauber) update \cite{Glauber1963}, implemented with a checkerboard (two-sublattice) sweep for cache-friendly local updates: For each site $i$ we draw $\sigma_i=+1$ with probability $P(\sigma_i{=}+1)=\{1+\exp[-2\beta(J\sum_{j\in nn(i)} \sigma_j+H_r)]\}^{-1}$, where $\beta=1/T_r$ and $H_r=\mu_a\,h_{\rm eff}(h,T_r)$ is the (temperature-dependent) field coupling of the replica. Replica exchanges between neighboring temperatures are attempted every five sweeps and accepted with the Metropolis criterion \cite{Metropolis1953} generalized to parameter-dependent Hamiltonians (replica exchange) \cite{Hukushima1996}: Swapping configurations $[s^{(r)},s^{(r+1)}]$ between $(T_r,H_r)$ and $(T_{r+1},H_{r+1})$ is accepted with probability $\min\{1,\exp(\Delta)\}$, where $\Delta=-\beta_r \mathcal{H}[s^{(r+1)};T_r,H_r]-\beta_{r+1}\mathcal{H}[s^{(r)};T_{r+1},H_{r+1}]+\beta_r \mathcal{H}[s^{(r)};T_r,H_r]+\beta_{r+1}\mathcal{H}[s^{(r+1)};T_{r+1},H_{r+1}]$. After discarding an initial thermalization stage of $n_\mathrm{therm}=20\,000$ sweeps, we measure the backbone magnetization $m_a=\langle \sigma_i\rangle$ for $n_\mathrm{measure}=100\,000$ sweeps (with convergence tested against $n_\mathrm{therm}=10\,000$ and $n_\mathrm{measure}=50\,000$)
 and estimate statistical uncertainties by binning measurements into blocks of 50 sweeps and taking the standard error over block means (blocking analysis) \cite{Flyvbjerg1989}.
}

\rrrred{Figure~\ref{Fig:2D_MC} depicts the $T$ dependence of $m_a(h,T)$ for the two cases: (i) $T_0^\mathrm{max}<T_c^*$ with $J=1$ and $T^*_c\simeq 2.269/k_\mathrm{B}$---the spin reversal takes place at $T_0(h)$, whose value decreases from $2.053/k_\mathrm{B}$ at $h=0.1$ [Fig.~\ref{Fig:2D_MC}(a)] to $1.451/k_\mathrm{B}$ at $h=1.5$ [Fig.~\ref{Fig:2D_MC}(b)]---and (ii) $T_0^\mathrm{max}>T_c^*$ with $J=0.88$ and $T^*_c \simeq 2.0/k_\mathrm{B}$---the spin reversal does not take place when $T_0(h)>T_c^*$ [Fig.~\ref{Fig:2D_MC}(c)] but occurs when $T_0(h)<T_c^*$ [Fig.~\ref{Fig:2D_MC}(d)].}

\rred{Note that $h_\mathrm{eff}$ is different from $h$ and the half-fire, half-ice--driven spin reversal is significantly different from spontaneous symmetry breaking. Spontaneous symmetry breaking takes place at finite $T_c$ for $h=0$; at $T_c$, $h\to0^+$ or $0^-$ can reverse the spins, but neither changing $T$ nor changing the magnitude of $h$ (not its direction) can reverse the spins. By contrast, the half-fire, half-ice--driven transition is absent for $h=0$; finite $T_0$ exists only for $h\ne 0$. Then at the finite $T_0$, $h_\mathrm{eff}\to 0^+$ or $0^-$ can reverse the spins, i.e., a slight change in $T$ or the magnitude of $h$ (without changing its direction) across the $T_0(h)$ curve can reverse the spins, as shown in Fig.~\ref{Fig:2D_MC}.}

\vspace{1cm}
\section{Discussion\label{SecDiscuss}}

The present results have immediate implications and inspire a series of subsequent studies, as outlined below.

\ignore{
We have not exactly solved the long-standing problem of the 2D and 3D Ising models in a magnetic field. However, by leveraging the key feature of $h_\mathrm{eff}=0$ in the site-decorated Ising models, we have exactly revealed the spin reversal transition at $T_0$ in the 2D and 3D Ising models under an external field. Furthermore, a reexamination of 2D and 3D bond-decorated models with ferromagnetic $J$ [Fig.~\ref{Fig:structure}(c)] also reveals a spin reversal transition (not shown), though it is less obvious.  These findings highlight the need for Monte Carlo simulations~\cite{Ising_FPGA} 
to accurately map the complete $h-T$ phase diagram of such decorated 2D and 3D Ising models \red{and compare it with Fig.~\ref{Fig:2D_MF} obtained in the hybrid-exact-mean-field theory}.
}

Traditionally, spin reversal in FMs and ferrimagnets is induced by applying a magnetic field in the opposite direction. In applications such as magnetic recording, overcoming the coercive force often requires increasing the local temperature close to the Curie temperature $T_c$, typically through laser illumination, so that a small applied field can trigger magnetization reversal~\cite{Kim_NM_22_review_ferri}. By contrast, the spin reversal mechanism presented here is particularly attractive, as it can be achieved through a slight change in temperature or magnetic field. This transition is driven by the hidden half-ice, half-fire state, in which the decorated \typeb sublattice provides a macroscopic degeneracy of $2^N$, resulting in an extremely sensitive and energy-efficient response of the \typea spins at $T_0$.
Notably, this response persists as $h\to 0$, particularly in 2D and 3D. The abrupt contrast in system behavior between $h=0$ and $h\to 0$ is rare and could serve as a method for measuring absolute zero magnetic field.

For the experimental realization of site-decorated models, mixed $d$-$f$ or $3d$-$5d$ compounds are promising candidates~\cite{Kim_NM_22_review_ferri,Ramirez_25_SmMn2Ge2}. In these systems, the \typea and \typeb spins correspond to $d$ and $4f$ elements, respectively, since the electronic wavefunctions of $d$ orbitals exhibit greater overlap than those of $4f$ orbitals. The weak interaction between $4f$ moments allows them to act as decorated spins, and the typically larger magnetic moments of $4f$ elements compared with 
$d$ ones naturally satisfy the condition $\mu_b> \mu_a$. A similar argument applies to $3d$-$5d$ compounds. Additionally, optical lattices~\cite{Bernien_Nature_17_Rydberg} and neural networks~\cite{ML_Hopfield_82,Schneidman2006}, both of which have been employed to study other Ising models, present alternative platforms for exploring site-decorated systems. 

\rrred{However, we caution that the extremely high anisotropy implied by Ising spins can only be expected in magnetic systems with strong spin-orbit coupling such as rare-earth ions or in a few exceptional cases involving transition-metal ions such as Co$^{2+}$ in highly distorted octahedral environments~\cite{Jongh_AiP_74_Ising_review,Stryjewski_AiP_77_Ising_review,Wolf_00_Ising_review}. The isotropic three-component Heisenberg spins  $\mathbf{S}_i=\{S_i^x,S_i^y,S_i^z\}$ are much more common in magnetic materials but also much more difficult to deal with---e.g., finite-$T$ phase transitions are prohibited in both 1D and 2D Heisenberg models with short-range interactions~\cite{Mermin_PRL_theorem}. The results of the Ising models are used to inspire the study of the corresponding Heisenberg models; indeed, the present results have motivated us to demonstrate the existence of UNPC in 2D site-decorated Heisenberg models where both the $a$ and $b$ spins are Heisenberg spins~\cite{yin_7_CMM_arXiv}.} 

\rrred{In the case where the $a$ spins are Ising spins described by Eq.~(\ref{ordinary}) but the $b$ spins are quantum Heisenberg spins described by $H_b=-\sum_{i}(J_{ab}\sigma_{i} S_{ib}^z+ h\mu_b S_{ib}^z)$, the $b$ spins can still be summed out exactly~\cite{Fisher_PR_59_Ising_transform}, reproducing Eqs.~(\ref{Z})--(\ref{heff}) with Eq.~(\ref{eq:boxedpm}) being replaced by
\begin{equation}
\boxed{\pm}=\frac{\sinh[(S_b+\tfrac12)\beta( h\mu_b \pm J_{ab} )]}{\sinh[\tfrac12\beta( h\mu_b \pm J_{ab} )]},    
\end{equation}
where $S_b$ is the spin value of $\mathbf{S}_{ib}$: $|\mathbf{S}_{ib}|=\sqrt{S_b(S_b+1)}$. For $S_b=1/2$, $\boxed{\pm}=2\cosh[S_b\beta (h\mu_b \pm J_{ab})]$, similar to Eq.~(\ref{eq:boxedpm}). This Ising-Heisenberg spin model also significantly enlarges the possibility of UNPC materialization in 1D.}

\begin{figure}[t]
    \begin{center}
        \subfigure[]{
\includegraphics[width=0.43\columnwidth,clip=true,angle=0]{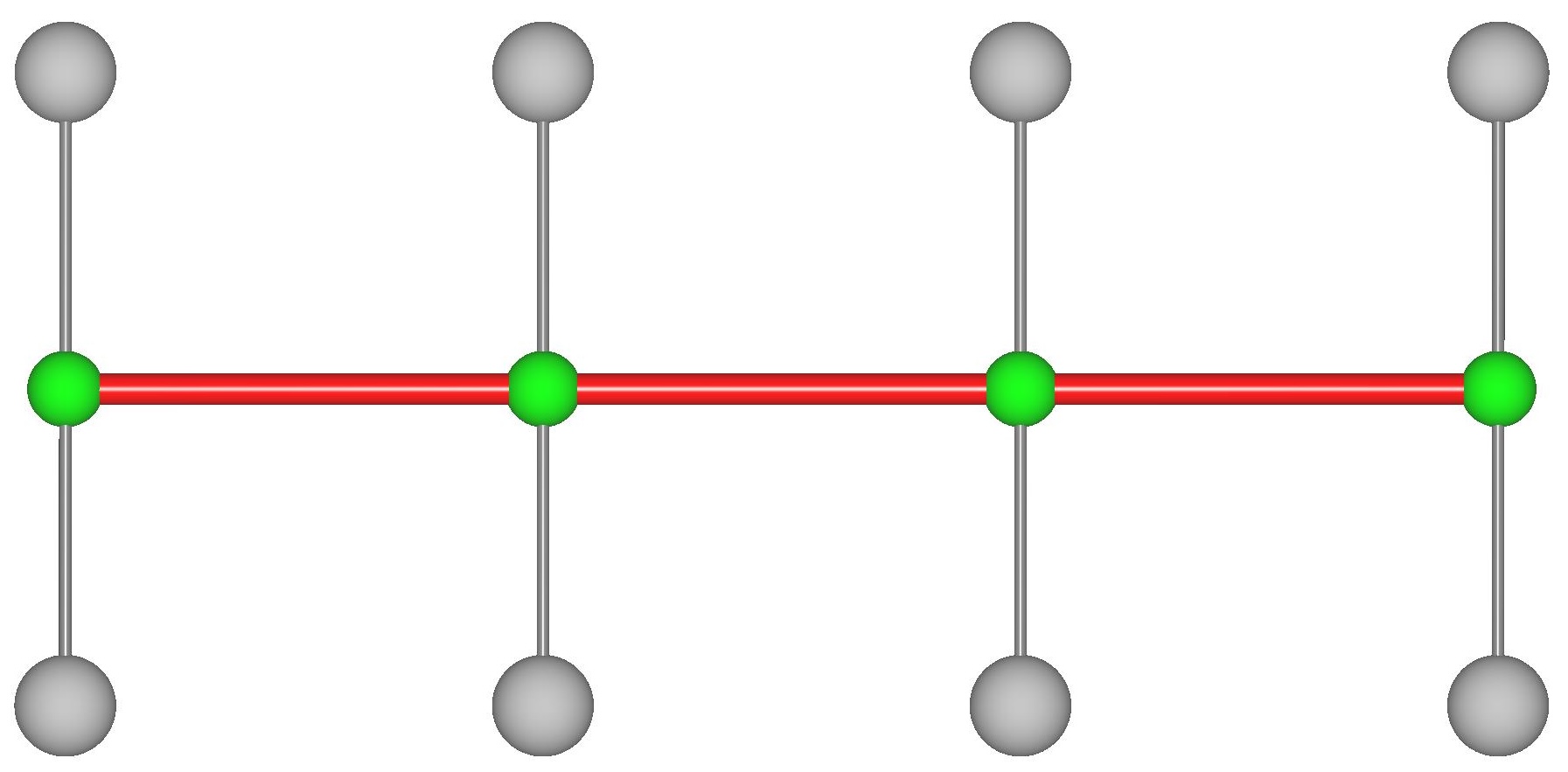}
        }
        \subfigure[]{
\includegraphics[width=0.48\columnwidth,clip=true,angle=0]{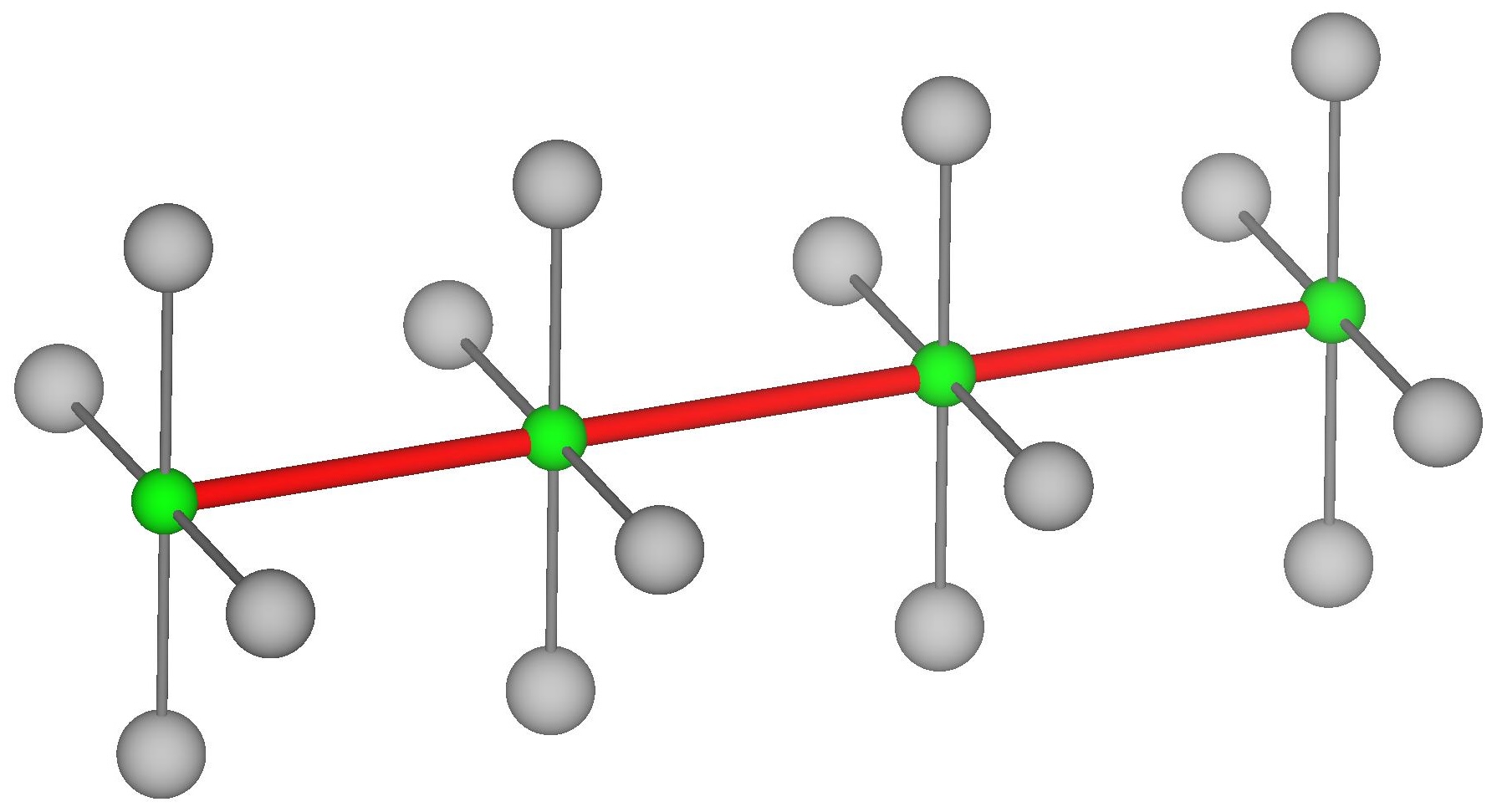}
        }
    \end{center}
\caption{$M$-side site decoration with (a) $M=2$ and (b) $M=4$, which effectively reduces the ratio of $\mu_a/\mu_b$ by a factor of $M$.}
\label{Fig:side}
\end{figure}

Another way to satisfy the condition
 $\mu_a /\mu_b < 1$ is through multiside ($M$-side) site decoration. Figures~\ref{Fig:structure}(b) and \ref{Fig:structure}(d) represent $M=1$. The examples of $M=2$ and $4$ are illustrated in Fig.~\ref{Fig:side}. Then, the effective magnetic field becomes
\begin{equation}
    h_\mathrm{eff}= h + \frac{M}{2\beta\mu_a}\left(\ln \boxedp- \ln \boxedm\right). \label{heffM}
\end{equation}
This results in the replacement of $\mu_a$ by $\mu_a/M$ in Eqs.~(\ref{eq:T0}) and (\ref{eq:T0max}), i.e., $\mu_a/\mu_b$ is effectively reduced by a factor of $M$. Consequently, $T_0$ can increase dramatically.

Finally, \rrred{since the site- and bond-decorated models are essentially equivalent in the large $J$ limit, the results of the site-decorated models are to inspire the study of the corresponding bond-decorated models, which are more difficult to deal with. In other words, the site-decorated model enables deeper understanding and faster identification of exotic phase-switching phenomena in any dimension, motivating more challenging studies with the bond-decorated model. In addition,} geometric compatibility of site and bond decorations significantly expands the potential for functional materials design. Moreover, advancing our understanding of the complexity and vast possibilities of the Ising model---a fundamental framework for cooperative phenomena across physical, biological, economic, and social systems---provides a crucial foundation for machine learning and AI applications in science.

\section{Summary}
In summary, site decoration has been introduced as an \red{alternative}  scheme for frustrated magnets \red{to the traditional bond decoration}, enabling \rrred{deeper understanding and} faster identification of exotic phase-switching phenomena in any dimension and yielding rare exact results for 2D and 3D systems in a finite magnetic field. \rred{The resulting unconventional frustration and physics are clarified by exactly mapping the 1D site-decorated Ising model in a magnetic field onto a more intricate 1D zero-field bond-decorated $J_1$–$J_2$ Ising model with conventional geometrical frustration.} These discoveries establish a paradigm for designing materials and devices with extreme sensitivity, where a hidden frustrated state---the half-ice, half-fire state---drives an abrupt spin reversal \rred{by a slight change in temperature or the magnitude of the external magnetic field (without changing its direction)} even under an infinitesimal applied field. In addition, the site-decorated model offers a rigorous test ground for AI reasoning and machine learning in science.

\begin{acknowledgments}
The author is grateful to Yabin Chen for stimulating discussions. 
Brookhaven National Laboratory was supported by U.S. Department of Energy,  Office of Basic Energy Sciences, Division of Materials Sciences and Engineering under Contract No. DE-SC0012704.
\end{acknowledgments}

\section*{DATA AVAILABILITY}

\rrrred{The data and code that support the findings of this article are openly available~\cite{Yin_Ising_IV_data,Yin_Ising_IV_github}.}

\begin{figure}[t]
    \begin{center}
\includegraphics[width=0.99\columnwidth,clip=true,angle=0]{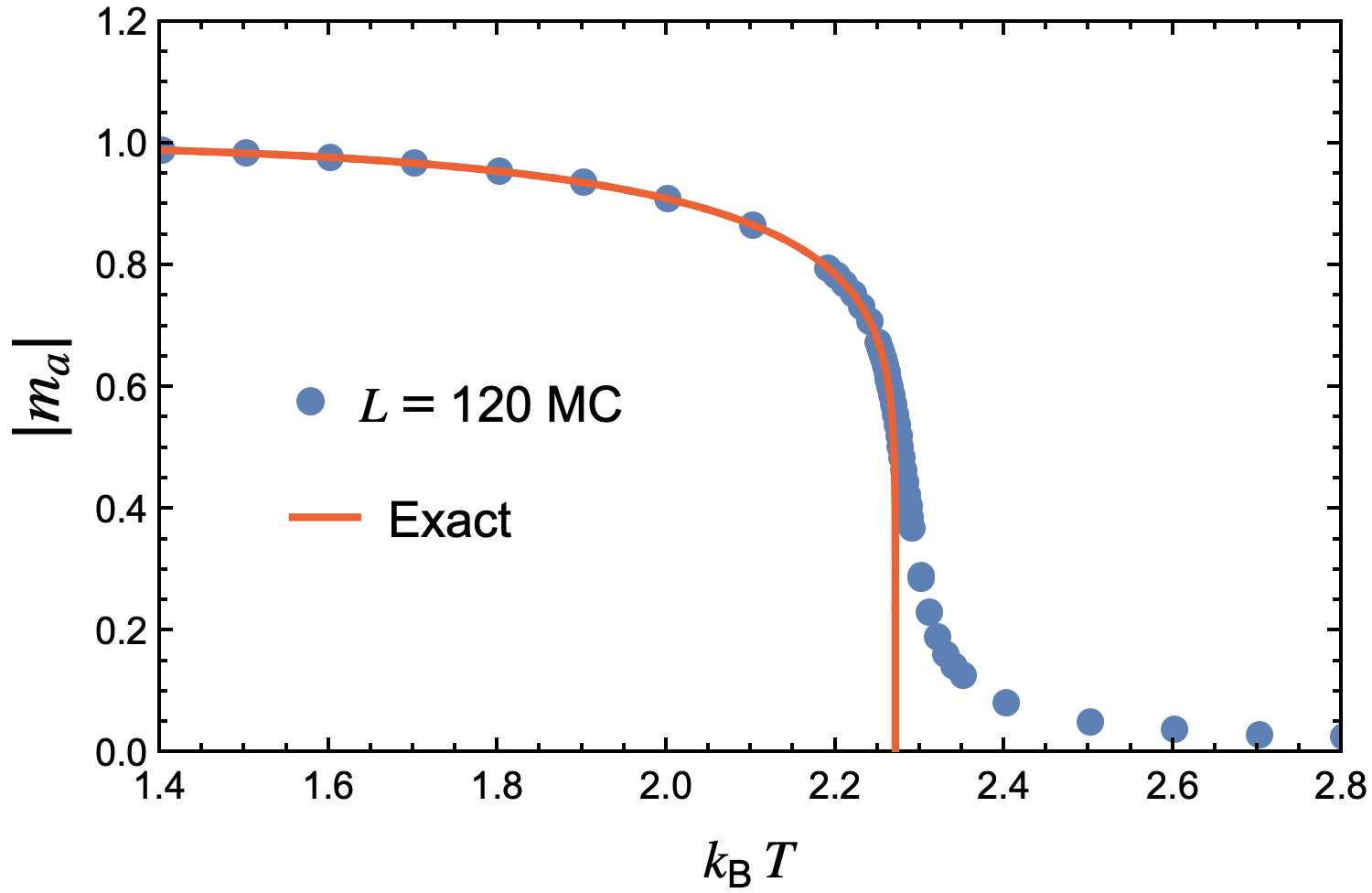}
    \end{center}
\caption{\rrrred{The $T$ dependence of zero-field $|m_a|$ in a standard Ising square lattice with $J=1$. The MC results for $L=120$ (blue dots) are compared with Onsager's exact results (red line).}}
\label{Fig:2D_MC_h0}
\end{figure}

\section*{Appendix A: AI-Assisted Discovery}

\hypertarget{AI}{The site-decorated model} offers a rigorous test ground for AI in science. During the 1000-Scientist AI Jam Session held on February 28, 2025, we tested the \texttt{OpenAI o3-mini-high} reasoning model by asking the AI to read the original version of this manuscript and redo the derivations. The AI did its own math to verify all our derivations. 

For example, it derived Eq.~(\ref{eq:mb}) using a theorem about probability, simpler than our previous derivation using $\langle b_i \rangle = -\frac{\partial f}{h\,\partial \mu_b}$. 

More impressively, it derived a much more elegant critical equation that determines $T_0$, Eq.~(\ref{eq:T0}), as elaborated below. 

\red{Initially, the critical equation was given by
\begin{equation}
    \exp(2\beta J_{ab}) =\frac{\sinh[\beta h(\mu_b - \mu_a)]}{\sinh[\beta h(\mu_b + \mu_a)]}, \label{eq:T0orig}
\end{equation}
which is the same as the critical equation for the minimal bond-decorated model except that the factor in front of $\beta J_{ab}$ is $4$, since each decorating spin is coupled to two backbone spins in bond decoration~\cite{Yin_Ising_III_PRL}.}

\red{The critical Eq.~(\ref{eq:T0}) derived by the AI is 
\begin{equation}
\tanh(-\beta J_{ab}) =\frac{\tanh(\beta h\mu_a)}{\tanh(\beta h\mu_b)}. \nonumber
\end{equation}
This equation is equivalent to Eq.~(\ref{eq:T0orig}). It elegantly makes the following features apparent:
\begin{enumerate}[label=(\arabic*)]
  \item $T_0$, the solution of the equation, is independent of $J$.
  \item $T_0$ is finite for $|\mu_a| < |\mu_b|$,  $\mu_a\mu_bJ_{ab}<0$, and $0<h<h_c$ or $-h_c<h<0$, where $h_c\equiv|J_{ab}/\mu_a|$.
  \item Finite $T_0$ exists in the zero-field limit $h\to 0$.
  \item $T_0$ continuously decreases to zero as $h$ increases to $h_c$.
\end{enumerate}
}

\red{We demonstrated that AI not only validated but also improved the analytic derivations in this paper. However, the AI first derived a lengthy expression for the critical equation. Guided by Eq.~(\ref{eq:T0orig}), we prompted the AI with the following question:
``When solving $T_0$, can you make $J_{ab}$ appear only on the left-hand side of the equation and $\mu_a$ and $\mu_b$ on the right-hand side?'' 
The AI reasoned for 25 s and produced the astonishingly elegant Eq.~(\ref{eq:T0}).}

\red{This successful human-AI interaction has inspired the further use of AI as scientific discoverer, e.g., in an AI-bootstrapped discovery (7d)~\cite{Yin_Potts_J1-J2_1D} and an AI-co-led discovery (5d)~\cite{Yin_Potts_UNPC}.}

\rred{In addition, for the resubmission of this manuscript, \rrrred{\texttt{OpenAI ChatGPT 5.2 Thinking} helped code the exact-mapping MC method in the C++ programming language (Sec.~\ref{hybrid}), which was validated by comparison with the exact result (Fig.~\ref{Fig:2D_MC_h0}) and the existing MC data for $h=0$~\cite{Ising_FPGA}---the performance of our code for $L=120$ sits between those for $L=128$ and $256$ in Ref.~\cite{Ising_FPGA}}. This uplifts the contribution of AI from AI-validated (2d) to AI-assisted (3d), according to a nine-level rating~\cite{Yin_Potts_UNPC}.}

\section*{Appendix B: Software} 

Wolfram Mathematica 14.2 was used to generate data for the exact solutions and plot Figs.~\ref{Fig:heff}--\ref{Fig:2D_MC} and \ref{Fig:2D_MC_h0}~\cite{Yin_Ising_IV_data}. 

VESTA 3.5.8~\cite{VESTA} was used to plot Figs.~\ref{Fig:structure} and \ref{Fig:side}. 

%


\end{document}